# The internal dynamics of very flattened normal galaxies
## Stellar distribution functions for NGC 4697 *

H. Dejonghe[1], V. De Bruyne[1]**, P. Vauterin[1]**, and W.W. Zeilinger[2,3]

[1] Sterrenkundig Observatorium, Universiteit Gent, Krijgslaan 281, B–9000 Gent, Belgium
[2] Institut für Astronomie, Türkenschanzstraße 17, A–1180 Wien, Austria
[3] Space Telescope - European Coordinating Facility, Karl Schwarzschild–Straße 2, D–85746 Garching bei München, Fed. Rep. Germany

**Abstract.**
Using the photometry and the kinematic data, part of which acquired with the NTT, we constructed a 2-integral and a 3-integral distribution function for NGC 4697. We detected a nuclear dust lane at $3.4''$ or 0.4 kpc from the centre.

A comparison of the Lucy-deprojection method and the multi-gaussian expansion method showed that the latter offers a number of advantages, among which a qualitatively better fit and better controllable convergence. Futhermore, we show that a 2-integral model cannot be excluded for as far as the Satoh parameter $k$ is allowed to vary. The 3-integral model is based on a Stäckel third integral. The original potential is retained where appropriate, in order to minimise errors due to the Stäckel approximation. On the whole, the 3-integral model produces more satisfying results than the 2-integral model, and is better in handling some peculiarities in the kinematic data. For both models we show detailed contour plots for all deprojected moments up to order 2. The mass-to-light ratio appears to be a fairly well determined $(M/L)_B = 4.8 h_{50}$, and is much better constrained than would be the case for spherical anisotropic models with comparable data.

**Key words:** Galaxies: elliptical and lenticular, individual: NGC 4697, kinematics and dynamics, structure.

## 1. Introduction

No really comprehensive model of a flattened galaxy has been made so far, which is a fit to photometric and kinematic data and goes as far as producing a distribution function. The main causes are (a) on observational side, there are very few objects with good kinematical data, i.e. data with good coverage of the 2D image, and reasonable errorbars and (b) models based on both 2 and 3 integral distribution functions are needed. In this paper, we will build such 2 and 3 integral models for the galaxy NGC 4697. This, in particular, means that we are fully using the strong (and obvious) constraint that the distribution function must be positive everywhere, in contrast with approaches solely from the Jeans equations. The main questions we want to address are:

- Can we rule out 2-integral models for NGC 4697?
- Are 3-integral models any better (regardless of the answer to the above question)?
- How constrained is the mass-to-light ratio?

We have chosen this particular galaxy for the following reasons:

- The photometry appears to be very symmetrical, which makes reasonable the assumption that the galaxy is in equilibrium.
- There is a fair amount of data available; the galaxy is well-studied, and was among the first cases where the observations proved to be discrepant with respect to the classic view of a rotationally flattened system (Bertola & Capaccioli 1975).
- NGC 4697 is very flattened; it may well be S0. In support of this view, Peletier et al. (1990) report a relatively strong disk. Moreover, the stellar rotation is substantial; the spheroidal part of NGC 4697 may very well be a large bulge.
- The galaxy is not remarkable in other wavelengths, which is a further indication that it is "normal".

There are two papers that consider dynamical models for NGC 4697. In Binney et al. (1990) the bulk of the kinematical data, which we also will use to large extent is presented. They use these to produce 2-integral models based on the Jeans equations. These are sufficient to determine $\sigma_\varpi^2 = \langle v_\varpi^2 \rangle = \langle v_z^2 \rangle = \sigma_z^2$ and $\langle v_\phi^2 \rangle$. The mean streaming $\langle v_\phi \rangle = \mu_\phi$ on the other hand may follow by introducing a parameter $k$ (Satoh 1980) such that

$$\sigma_\phi^2 = \langle v_\phi^2 \rangle - \mu_\phi^2 = k^2 \sigma_\varpi^2 + (1-k^2)\langle v_\phi^2 \rangle, \tag{1}$$

or

$$k = \frac{\mu_\phi}{\sqrt{\langle v_\phi^2 \rangle - \sigma_\varpi^2}}. \tag{2}$$

---

Physically, $k = 1$ means an isotropic velocity dispersion tensor. Their models turned out to be such that $k < 1$, or $\sigma_\phi > \sigma_\varpi$, since in general the excess of rotational energy leads to $\langle v_\phi^2 \rangle > \sigma_\varpi^2$. Binney *et al.* conclude that NGC 4697 is only marginally fit by such models, with the rotation consistently high even for $k \approx 0.7$. One may remark however, that there is no reason to assume that $k$ is a constant throughout the galaxy, apart from the obvious operational necessity to do so in the modeling using the Jeans equations.

Merrifield (1991) also considers the Jeans equations, but works more in the observational domain. He uses integral forms for the projected quantities, very much in the same style as in the pioneering work of Binney & Mamon (1982). Interestingly, he develops a global diagnostic to see whether $k$ and/or the mass-to-light ratio $\Upsilon$, or a combination of both, should vary. Not surprisingly, he finds this to be the case, but infers also that there is a detectable influence of a third integral. The amount of anisotropy resulting from that he deems modest however, at least in the case of maximum tilt of the velocity ellipsoid.

In the next section, we will discuss the available data. Section 3 concerns the determination of the mass density and the potential, which are prerequisites for any dynamical model. The modeling is done using Quadratic Programming (QP). The essentials of it, in a stellar dynamical context, are discussed in Dejonghe (1989) or Batsleer & Dejonghe (1995). In Section 4 we construct 2-integral models, and in section 5 3-integral models are considered. The results and conclusions follow in sections 6 and 7.

## 2. Observations and data reduction

**Fig. 1.** NGC 4697: a co-addition of two PC-I images after applying 10 iterations of the Richardson-Lucy algorithm.

The observations were carried out on 17-18 May 1991 at the ESO NTT with the Red Channel of EMMI using as detector a FA 2048 × 2048 CCD (ESO CCD #24) with a pixel size of 15µm. The scale was 0.347"/pixel.

Long-slit spectra of NGC4697 were obtained at various position angles as specified in Table 1. The dispersion of the grating was 28 Å/mm in the efffective wavelength region of $\lambda\lambda 5030 - 5650$ Å. The slit width of 0.7" yielded a spectral resolution of $\approx 20$ km/s on the detector, as determined from the average line widths of the Argon-lamp calibration spectra. The CCD spectra were reduced using the ESO-MIDAS program package, following standard procedures. All CCD frames were bias subtracted, corrected for dark current, flat fielded and the cosmic ray events were cleaned. The slit transfer function was corrected with sky spectrum obtained at dawn. The wavelength calibration of the spectra was performed using the mean of two Argon arc spectra obtained before and after each science spectrum in order to take possible instrumental distortions during the science exposure. The wavelength calibration was carried out by means of fitting a third order polynomial to each spectral row separately. The accuracy of the fits was usually better than 0.1 pixel. The galaxy spectra were sky subtracted using the areas uncontaminated by galaxy light near the edges of the slit. The galaxy spectra were then rebinned spatially in order to obtain spectral rows having $S/N \geq 15$. A modified version of Fourier Quotient technique (Bertola et al. 1984) was used to derive radial velocities and velocity dispersions as a function of galaxy radius. A spectrum of the K-giant star HR4699 taken with the same instrumental setup yielded the velocity and line-broadening template.

A $120^s$ V band image of NGC 4697 was obtained in the night of 17/18-May-1991. The FWHM of the seeing was about 1" as determined from stellar images in the CCD frame. The correction for bias, dark current, flat field was performed in a similar procedure as for the spectra. The aperture photometry compilation of Longo & de Vaucouleurs (1988) was used for the V band surface brightness calibration. In order to study the morphology of the nuclear region of NGC 4697 two F555W *Planetary Camera-1* images of 500 sec exposure time each obtained from the HST Science Data Archive were analyzed. The image scale is 0.0445"/pixel. The two frames were aligned using the bright galaxy centre as reference point. The accuracy of the alignment is about 1 pixel. The cosmic ray events were removed applying standard procedures. The images were deconvolved by applying 10 iterations of the FFT Richardson-Lucy image restauration algorithm (Hook & Lucy 1992), maximizing the likelihood each step, and co-adding them by convolution of a "core" PSF obtained with the TinyTim software package (Kirst 1992). A description of this procedure is given by Zeilinger (1994). The effective resolution of the restored image is about 0.2", as determined from several star-like sources in the galaxy image. The most prominent feature is, in addition to the large-scale stellar disk, a nuclear dust lane. The dust lane has a smooth ring-like appearance, suggesting an almost edge-on view. The dust lane surrounds an unresolved central source and may indicate the presence of a gas disk having linear diameter of $\simeq 800$ pc. Such central structures are nowadays often detected in elliptical galaxies (e.g. Jaffe *et al.*1994). A FITS file of the restored HST image is available via anonymous ftp on doradus.ast.univie.ac.at (131.130.36.64) in /pub/hst/N4697.fits.

The following table (2) summarizes the new data used for the modeling. The other kinematical data were taken from Binney *et al.* (1990). In table (3) we present the data at 5" offset parellel to major axis and at 45 deg intermediate axis. We have not used this data for the modeling, with the intention to obtain results that are comparable with those of Binney *et al.*.

| P.A. | exposure time | comments |
|---|---|---|
| 63° | 3600+5400$^s$ | major axis |
| 63° | 3600+5400$^s$ | 5″ offset ∥ major axis |
| 108° | 5400$^s$ | 45° intermediate axis |

**Table 1.** The observing log.

| $x''$ | $\langle v_p \rangle$ | $\delta \langle v_p \rangle$ | $\sigma_p$ | $\delta \sigma_p$ |
|---|---|---|---|---|
| 0.0 | 0.0 | 10 | 170.0 | 10 |
| 1.0 | 67.7 | 10 | 170.9 | 13 |
| 2.1 | 102.1 | 10 | 172.9 | 14 |
| 3.1 | 108.4 | 10 | 177.2 | 15 |
| 4.2 | 113.2 | 10 | 180.0 | 15 |
| 5.2 | 115.3 | 10 | 184.6 | 15 |
| 6.6 | 115.3 | 10 | 189.7 | 15 |
| 8.3 | 114.6 | 10 | 197.1 | 15 |
| 10.0 | 113.2 | 10 | 210.0 | 16 |
| 13.5 | 112.9 | 10 | 200.6 | 21 |
| 18.0 | 110.1 | 10 | 186.0 | 23 |
| 25.0 | 109.4 | 10 | 176.9 | 28 |
| 34.0 | 105.7 | 10 | 170.5 | 32 |
| 43.0 | 102.7 | 10 | 165.0 | 36 |
| 50.0 | 100.4 | 10 | 167.9 | 38 |

**Table 2.** The new kinematical data along the apparent major axis, as used in the modeling, all velocities are in km/s.

**Fig. 2.** The axisymmetrised V-image of NGC 4697 (NTT). Contours are labeled in magnitudes. The effective radius (half light) is about 120″. Note the presence of disky isophotes.

## 3. The potential

Fig. 2 shows the calibrated visual image of N4697. It is a 601 × 416 CCD frame, from which foreground stars were removed. The isophotes look very regular, and point to a flattening of .55 (hence E4.5) at 50″. The presence of the disk (Carter 1987, Peletier *et al.* 1990) is clearly visible. The image is also very symmetric, in the sense that, where available, $I(x,y) = I(-x,y) = I(x,-y) = I(-x,-y)$. We have forced that symmetry by replacing corresponding pixels by their mean value. The final result is almost indistinguishable from the original, as shown by the consistently low values in the histogram of the symmetry deviations (figure 3).

| $x''$ | $\langle v_p \rangle$ | $\delta \langle v_p \rangle$ | $\sigma_p$ | $\delta \sigma_p$ |
|---|---|---|---|---|
| 5″ | | | | |
| 0.2 | 14.3 | 8 | 180.5 | 5 |
| 0.9 | 61.4 | 8 | 176.3 | 5 |
| 1.9 | 95.5 | 9 | 158.9 | 6 |
| 3.6 | 95.5 | 12 | 163.9 | 7 |
| 5.4 | 81.5 | 16 | 171.6 | 7 |
| 6.8 | 108.0 | 15 | 171.9 | 7 |
| 8.2 | 113.3 | 20 | 185.0 | 8 |
| 10.1 | 108.5 | 20 | 198.7 | 8 |
| 12.4 | 96.9 | 19 | 178.5 | 8 |
| 18.9 | 109.4 | 10 | 173.4 | 6 |
| 26.1 | 107.3 | 14 | 171.5 | 7 |
| 33.3 | 110.2 | 20 | 159.3 | 8 |
| 37.4 | 97.6 | 26 | 160.8 | 10 |
| 45° | | | | |
| 1.6 | 41.1 | 8 | 173.1 | 5 |
| 2.6 | 52.8 | 10 | 169.6 | 6 |
| 4.0 | 42.0 | 12 | 183.9 | 6 |
| 4.7 | 54.3 | 1 | 159.9 | 6 |
| 6.4 | 48.8 | 12 | 169.9 | 7 |
| 7.8 | 68.3 | 15 | 164.6 | 7 |
| 8.5 | 45.8 | 13 | 157.7 | 7 |
| 10.6 | 18.9 | 16 | 166.4 | 8 |
| 13.2 | 34.7 | 17 | 170 | 8 |
| 16.8 | 35.8 | 25 | 171.6 | 9 |
| 19.5 | 41.1 | 17 | 170.0 | 8 |
| 23.3 | 69.0 | 21 | 166.1 | 9 |
| 38.3 | 43.7 | 35 | 158.9 | 16 |

**Fig. 3.** The histogram of the deviations from axisymmetry in the V-image of NGC4697 (NTT). Abscis is expressed in promille of the surface brightness, which amounts to .0025 mag

**Table 3.** The kinematical data at 5" offset parallel to major axis and at 45° internmediate axis. All velocities are in km/s.

In order to obtain a luminous mass density (LMD) based on these data, we need to deproject the photometry. This necessitates a choice for the inclination $i$, and a distance, which we take to be 24 Mpc. There are two main deprojection procedures available from the literature. We have applied both techniques, and discuss the results below.

### 3.1. The inclination and the dust lane

Because of the presence of the dust lane, visible in the NTT image but very clearly visible in the HST image, we have a good handle on the inclination. If we assume that the dust lane has an intrinsically circular shape, and is located in the equatorial plane, we can easily deduce an inclination by ellipse fitting. In fact, we obtained a very good fit, corroborating our assumption of a circular ring, and we obtain $i = 78° \pm 5°$. We will adopt $i = 80°$ as a very workable value. The presence of the dust lane at about $3''$ further amply demonstrates a detectable interstellar medium for this galaxy. In fact, Goudfrooij et al. (1994) commented on the likeliness of cold gas because of high IRAS flux densities, and a very red nucleus. They detected ionized gas, extending over $35''$.

### 3.2. Lucy deprojection

Binney et al. (1990) use an iterative Bayesian procedure (Lucy 1974). In order to implement it, one needs to interpolate a 2-dimensional function given on a grid. We chose an elliptically polar grid, $b/a = .6$, with rays that need not have equally spaced angles, and bicubic splines. The semimajor axes are linearly spaced at small galactocentric distances, and logarithmically spaced at large radii. The initial guess for the luminous mass density was taken to be the best-fit Hubble model. Typically 7-8 iterations could be performed before the LMD did not improve anymore. For every inclination considered, we obtain a LMD on an elliptically polar grid, again to be evaluated using bicubic splines. The quality of the inversion can be judged by projecting this LMD again, and comparing it with the original photometry. This is done in figure 4, where the statistics of the deviations are shown. Clearly $i = 80°$ yields the smallest errors and a well-behaved histogram.

**Fig. 4.** Histograms of the deviations between the projected LMD and the axisymmetrical photometry, for different inclinations (Lucy algorithm) and for the gaussian deprojection (all inclinations).

This result probably should not be overinterpreted. Certainly, it is gratifying that also here inclinations of the order of $i = 80°$ stand out as a good choice. A Bayesian inference scheme is based on statistical principles, and yields a distribution that is often characterized as "most plausible". However, the solution is a functional of the initial guess (in our case a Hubble profile), even more so in any practical implementation where one stops the procedure after a relatively small number of iterations. Only if $i = 90°$ is the axisymmetric deprojection unique, as is well known, and the indeterminacy exacerbates strongly towards smaller inclinations, as one can easily verify in the case $i = 0°$. Therefore, if the problem has multiple solutions, there is no reason that, in case of convergence, Lucy's scheme arrives at the same solution, independent of the initial guess. The good performance of the case $i = 80°$ must probably be somewhat qualified in this sense.

### 3.3. Gaussian deprojection

During the last years, the so-called multi-gaussian expansion method has surfaced as a valuable alternative to Lucy's (Bendinelli 1991, Monnet et al. 1992, Emsellem et al. 1993). It has the advantage that, if the PSF (Point Spread Function) is known and modeled by a set of gaussians as well, the seeing-deconvolved telescope image is fairly easily obtained. In our case however, this advantage does not weigh in very heavily, since we are not concerned with dynamical models for the very centre of NGC 4697. Further advantages include that, within the approximations, the deprojection of the gaussian decomposition is exact, analytical and smooth. A possible drawback may be that peculiarities in the photometry may not always be easy to fit with gaussians.

In order to obtain a deprojected axisymmetric LMD, one first fits the observed surface brightness by a series of gaussian functions with common centre and symmetry axes. The parameters are the intensity $I$ of the component, the dispersion $\sigma$ and the flatness $q$ of the two-dimensional gaussian component (Emsellem et al. 1993).

Let the photometry be specified on a grid of points $(x_i, y_i)$ by $I(x_i, y_i)$, $i = 1...n$, one then minimizes

$$\sum_{i=1}^{n} \left( I(x_i, y_i) - \sum_{k=0}^{m} I_k(x_i, y_i) \right)^2, \qquad (3)$$

by varying the $3m$ coefficients $I_k$, $\sigma_k$ and $q_k$. In principle, this is a non-linear fitting problem (Monnet et al. 1992), but we used a more indirect approach by discretising the values of the parameters $\sigma_k$ and $q_k$, in order to set up a library of components. The fit then is an iterative procedure. The first iteration consists in selecting the component that yields the best fit. Every subsequent iteration will select a best component, which is then added to the set. In this way, a simple linear fit is sufficient at each stage of the calculation.

Again assuming an inclination $i$, an axisymmetric three-dimensional LMD is projected on a two-dimensional gaussian function. One thus can associate a three-dimensional deprojected gaussian component with each two-dimensional fitting component.

Clearly, the gaussian deprojection can be developed in a more general frame in order to model multiple centres and isophote twists (Emsellem et al. 1993). Since the photometry of NGC 4697 is very regular, these complications do not concern us here.

Typically the gaussian fit was performed with a modest number of components, ranging from 12 to 18. Fig. (4) also contains the statistics of the deviations with the photometry for one of our gaussian expansions. The relevant histogram is the same for all inclinations considered, since in this case the library only contained components that were possible for all of them. Hence, this particular fit was independent of the inclination. Clearly, the gaussian fit is of comparable quality to the Lucy deconvolution for $i = 80°$, but distinctly better in all other cases. Also, at faint intensities, the Lucy fit suffers from grid effects to some extent (in particular on the short axis, where the elliptically polar grid was less dense), which are obviously absent in the gaussian scheme (by design). This can be seen in fig. (5), which also shows that both methods give very similar results. The intensities are defined as $2.512^{-V}$, with $V$ the surface brightness in magnitude per square arcsec.

**Fig. 5.** A comparison between the Lucy deprojection for $i = 80°$ and the gaussian deprojection. The upper curves are the deprojected intensities along the major axis, the lower curves are intensities along the minor axis. Full curves represent the Lucy deprojection, dot-dashes are for the gaussian decomposition.

**Fig. 6.** Contours for the LMD of NGC 4697 assuming $i = 70°$ (thin countours) and $i = 90°$ (thick contours) for the gaussian deprojection. The values for the contours can be derived from the previous figure. Note the presence of a disk in the $i = 70°$ deprojection, which is virtually absent in the $i = 90°$ deprojection. Clearly, the contours for the $i = 80°$ case which we use for the modeling are intermediate between the cases shown.

Fig. (6) shows contours of the LMD obtained by the gaussian decomposition for $i = 90°$ (thick contours) and $i = 70°$ (thin contours). As can be expected, the $i = 70°$ LMD needs to be flatter and more disky if it is to reproduce the same photometry. By and large though, both LMDs look very similar, and it is clear that deprojection alone is a poor diagnostic for inclination. The case $i = 80°$ is intermediate between the 2 cases, but is omitted here for clarity. Roughly, 10 kpc corresponds to $100''$, and therefore the LMD outside 10 kpc is less constrained by the data. At 5 kpc on the major axis we measure a flattening of $b/a = .54$.

One may argue whether not all deprojected mass densities should have shown a disk component. The gaussian decomposition clearly shows however that identical (projected) photometry may or may not produce a disk in deprojection, depending on the viewing angle. On the other hand, a disky deprojected mass density may very well produce a projected density that is as good a fit as the one we used here. Therefore, we will not assign large weight to the presence of the disk in this paper. The disentanglement of disk and bulge kinematics in the case of NGC 4697 is completely beyond the goal we set ourselves. At this point, we leave the issue open whether our dynamical models represent the entire galaxy in the region considered, or only the bulge of it.

Finally, as a matter of reference, table (4) provides a few numerical values for the LMD as computed from the gaussian decomposition, in units of $M_\odot/\text{pc}^3$, assuming a constant mass-to-light ratio $\Upsilon_V = 3.5$.

| $\varpi$ | $z = 0$ | $z = 1$ | $z = 2$ |
|---|---|---|---|
| 0.0 | 2.229 | -0.013 | -0.681 |
| 1.0 | 0.509 | -0.149 | -0.724 |
| 2.0 | -0.141 | -0.419 | -0.830 |
| 3.0 | -0.529 | -0.669 | -0.954 |
| 4.0 | -0.769 | -0.864 | -1.081 |
| 5.0 | -0.959 | -1.035 | -1.215 |
| 6.0 | -1.144 | -1.208 | -1.356 |
| 7.0 | -1.332 | -1.382 | -1.499 |
| 8.0 | -1.512 | -1.549 | -1.638 |
| 9.0 | -1.676 | -1.703 | -1.770 |
| 10.0 | -1.824 | -1.843 | -1.895 |

**Table 4.** The logarithm of the LMD (in $M_\odot/\text{pc}^3$) for the $i = 80°$ gaussian deprojection. An overall $(M/L)_V = 3.5$ was assumed. The coordinates $\varpi$ and $z$ are expressed in kpc.

### 3.4. The potential

Once the LMD is known, we can calculate the potential, if the mass-to-light function $\Upsilon(\varpi, z)$ is given. As above, we will assume, for better or worse, that it is a constant. Its precise value does not matter here now, since we will normalize the binding potential $\psi(\varpi, z)$ such that $\psi(0, 0) = 1$. The total mass inside the last kinematic data point will be used to scale the kinematic data.

The calculation of the potential is straightforward in principle. It suffices to decompose the LMD in spherical harmonics

$$\mathrm{LMD}(r,\theta) = \sum_{2k=0}^{K} \frac{4k+1}{2} \mathrm{LMD}_k(r) P_{2k}(\cos\theta), \quad (4)$$

with

$$\mathrm{LMD}_k(r) = 2\int_0^1 \mathrm{LMD}(r,\theta) P_{2k}(\cos\theta)\, d(\cos\theta). \quad (5)$$

In these formulae, $(r, \theta)$ are spherical coordinates, the $P_k$ are Legendre polynomials, and there is the implicit assumption that $\mathrm{LMD}(r, z) = \mathrm{LMD}(r, -z)$. In practice, very good results are obtained for $K = 7$. The potential then follows from

$$\psi(r,\theta) = \sum_{2k=0}^{K} \psi_k(r) P_{2k}(\cos\theta), \quad (6)$$

with

$$\psi_k(r) = 2\pi G \left[ r^{-2k-1} \int_0^r \mathrm{LMD}_k(r')(r')^{2k+2} dr' \right.$$
$$\left. + r^{2k} \int_r^{+\infty} \mathrm{LMD}_k(r')(r')^{-2k+1} dr' \right]. \quad (7)$$

### 3.5. The Stäckel fit

As mentioned in the introduction, one of our goals is to explore the presence of a third integral in the observed kinematics. As is well known, in general no such integral exists in an axisymmetric potential. Moreover, the operational advantage of an analytic, and therefore approximate, integral is such, that we will fit the constant $\Upsilon$ potential with a Stäckel one, of the form

$$\psi(\varpi, z) = \frac{(\lambda+\gamma)G(\lambda) - (\nu+\gamma)G(\nu)}{\lambda - \nu}, \quad (8)$$

where $(\lambda, \nu)$ are the spheroidal coordinates of a point $(\varpi, z)$, i.e. the roots for $\tau$ of

$$\frac{\varpi^2}{\tau+\alpha} + \frac{z^2}{\tau+\gamma} = 1. \quad (9)$$

The method to accomplish this was originally devised by De Zeeuw & Lynden-Bell (1985), and implemented by Dejonghe & De Zeeuw (1988). It essentially takes advantage of the special form (8) by averaging $(\lambda-\nu)\psi(\varpi,z)$ once over $\nu$ and once over $\lambda$, with suitable weight functions. This of course presumes an assumption for the focal distance $\Delta = \sqrt{\gamma-\alpha}$. One can easily see that this yields $(\tau+\gamma)G(\tau)$, once for $-\gamma \le \nu \le -\alpha$ and once for $-\alpha \le \lambda$. The procedure then is repeated for different $\Delta$, and the best choice, i.e. the $\Delta$ and its Stäckel potential with the smallest deviations from the original potential, is finally retained.

In Figure (7) both are compared. The best fit Stäckel potential has a focal distance of .5 kpc, and the peak error nowhere exceeds 5% in the frame of the figure. Obviously, the Stäckel version is rounder than the original. This is a well-known property of Stäckel potentials, which has to do with the fact that the ellipses of constant $\lambda$ are confocal, and whence become rapidly round at larger distances from the centre. This is not necessarily a problem, though, since we have no idea whether the assumption of a constant mass-to-light ratio is a valid one. If not justified ultimately, the dark matter will probably be distributed in a distribution which is rounder than that of the stars we see, if our experience with spiral galaxies is of any guidance here. It is important to note, however, that we will never calculate the potential with (8), but that we will continue to use (6) for that purpose. Only in the expression of the third integral will we use the function $G(\tau)$ as defined in (8). Therefore, our models are of the class that use an approximate third integral (e.g. Binney & Petrou 1985).

**Fig. 7.** Contours for the potential of a constant $\Upsilon$ mass model for NGC 4697 (thick lines), compared with contours for the Stäckel fit (thin lines). The line spacing is 0.1 for both line types.

An alternative Stäckel fit can be obtained by using the property that a Stäckel mass model and potential is completely determined by their values on the $z$-axis (de Zeeuw 1985b). Hence, the function $G(\tau)$ can be expressed directly in terms of the potential along the z-axis:

$$G(\tau) = \frac{(\tau+\alpha)\psi(0,\sqrt{\tau+\gamma}) + (\gamma-\alpha)\psi(0,0)}{\tau+\gamma}. \quad (10)$$

This produces another Stäckelfit to the potential, by simply inserting (10) in (8). Obviously, this fit is perfect on the $z$-axis, but, not surprisingly, worse in the equatorial plane.

### 4. Two-integral models

The expressions for the distribution functions that are used for modeling the galaxy are constructed following the quadratic programming method as described by Dejonghe (1989). We consider the distribution function to be a sum of basis distribution functions $F_i$,

$$F(I) = \sum_i c_i F_i(I). \quad (11)$$

Observations of the galaxy result in a number $L$ of observed moments $\mu_l^\star(r_l, v_l)$, $l = 1, \ldots, L$, with errors $\delta\mu_l^\star(r_l, v_l)$. The only requirement is that these moments should be linear in the distribution function. Subsequently, the corresponding functions $\mu_{l,i}(r_l, v_l)$, $l = 1, \ldots, L$ are calculated for each basis distribution function $F_i$, so that their linear combination is to be equal to the observed function. Therefore one constructs the variable

$$\chi^2 = \sum_l w_l \left[ \mu_l^\star(r_l, v_l) - \sum_i c_i \mu_{l,i}(r_l, v_l) \right]^2, \quad (12)$$

which is a quadratic funtion of the coefficients $c_i$. The constants $w_l$ are used for normalisation and for tuning, or can, in special cases, be used to give the $\chi^2$ its usual statistical meaning. The minimization of this variable leads to a linear set of equations, the solutions of which are the coefficients $c_i$ for the basis functions in the distribution function $F(I)$. However, such a procedure would not take into account the requirement of positivity for $F(I)$. This we do by testing $F(I)$ on a grid in integral space, which yields additional linear constraints. Our original minimization problem hence becomes a quadratic programming problem.

In this section we consider functions $F(E, L_z)$, with $E = \psi(\varpi, z) - v^2/2$ and $L_z = \varpi v_\phi$. Two-integral models have the advantage that the even part is determined by the mass density, which, given the inclination of NGC 4697, is well known. The inversion is in principle very unstable (Dejonghe 1986, Hunter & Qian 1993), but the requirement of positivity probably renders the inversion much more stable (Batsleer & Dejonghe 1995). Despite this theorem, there is no point in fitting only the mass density, and leave the kinematics as a check, since this would put undue emphasis on the photometry. Moreover, this would also leave open the determination of the odd part of the distribution function, which must follow from the observed mean streaming.

Therefore, we include all data, i.e. photometry $\rho_p(x_i, y_i)$, projected fluxes $\rho_p(x_i, y_i)\langle v_p\rangle(x_i, y_i)$ and projected pressures $\rho_p(x_i, y_i)\langle v_p^2\rangle(x_i, y_i)$ in the $\chi^2$ variable. The projected velocity dispersion is given by $\sigma_p^2 = \langle v_p^2\rangle - \langle v_p\rangle^2$. With this choice of the observables, the $\chi^2$ has the form

$$\begin{aligned}\chi^2 &= w_{\mathrm{md}}\sum_i\left[\frac{\rho_p^*(x_i,y_i)-\rho_p(x_i,y_i)}{\delta\rho_p(x_i,y_i)}\right]^2\\&+ w_{\mathrm{mean}}\sum_i\left[\frac{\rho_p^*\langle v_p\rangle^*(x_i,y_i)-\rho_p\langle v_p\rangle(x_i,y_i)}{\delta\rho_p\langle v_p\rangle(x_i,y_i)}\right]^2\\&+ w_{\mathrm{disp}}\sum_i\left[\frac{\rho_p^*\langle v_p^2\rangle^*(x_i,y_i)-\rho_p\langle v_p^2\rangle(x_i,y_i)}{\delta\rho_p\langle v_p^2\rangle(x_i,y_i)}\right]^2.\end{aligned} \quad (13)$$

The 3 weights $w_{\mathrm{md}}$, $w_{\mathrm{mean}}$ and $w_{\mathrm{disp}}$ assigned to each of the three classes of data are arbitrary, and are chosen such that the obtained fit is optimal, as defined by simple inspection. Inside each class however, the weights for the individual data points scale properly according to the observational errors. Obviously, it is meaningless to discuss the absolute value of the $\chi^2$, since it is defined up to a multiplicative factor.

The components we use are of Fricke type:

$$\begin{aligned}F_{pq}^\pm &= E^p L_z^{2q} 2^{-q} \quad \pm L_z \geq 0\\&= 0 \qquad\qquad \pm L_z < 0,\end{aligned} \quad (14)$$

yielding mass densities

$$\rho_{pq}^\pm = \frac{\sqrt{2}\pi\Gamma(q+\tfrac{1}{2})\Gamma(p+1)}{\Gamma(p+q+\tfrac{5}{2})}\psi^{p+q+3/2}\varpi^{2q}. \quad (15)$$

All moments are given by

$$\left(\rho\langle v_\varpi^{2i} v_\phi^m v_z^{2j}\rangle\right)_{pq}^\pm = (\pm)^m g_{pq}^{i,m,j}\psi^{p+q+3/2+i+j+m/2}\varpi^{2q}, \quad (16)$$

with

$$g_{pq}^{i,m,j} = \frac{\Gamma(i+\tfrac{1}{2})\Gamma(j+\tfrac{1}{2})\Gamma(q+\tfrac{m+1}{2})2^{i+j+\tfrac{m+1}{2}}\Gamma(p+1)}{\Gamma(p+q+i+j+\tfrac{m}{2}+\tfrac{5}{2})}. \quad (17)$$

This choice of Fricke components (14) is motivated by the fact that all moments can be calculated analytically, and numerical integration is only necessary when integrating projected moments through the galaxy.

It is useful to consider also $F_{pq}^e = F_{pq}^+ + F_{pq}^-$, which are even components. None of these single components are even crude approximations to the data, and therefore we will not discuss them in any detail. However, it is useful to know that the larger $p$, the more concentrated the component. For $q = 0$ the components are stratified along the equipotential surfaces, and for $q > 0$ the components are tori. The "hole" in such a torus becomes larger with increasing $q$. For $0 < q < 1$ there is a cusp in the mass density. The ranges of the parameters $p$ and $q$ that were used to choose components from are given in table (5).

In practice, the odd part of the distribution function is taken into account by considering the even components (in which there are equal numbers of stars rotating in either sense), and one of the sets in (14) (we chose negative angular momentum, but that of course is arbitrary). The fit was first performed on the even set, taking into account only the data $\rho_p(x_i, y_i)$ and $\rho_p\langle v_p^2\rangle(x_i, y_i)$. To the resulting set of even components were then added the corresponding rotating components, in order to obtain the best mix reproducing also $\rho_p\langle v_p\rangle(x_i, y_i)$. Hence, the distribution function has the form

$$F(E, L_z) = \sum_{p,q}(c_{pq}^e F_{pq}^e(E, L_z) + c_{pq}^- F_{pq}^-(E, L_z)). \quad (18)$$

It is essential that one also allows for negative coefficients $c_{pq}^e$ and $c_{pq}^-$. The final 2-integral distribution functionis composed out of 56 components. We used $w_{\mathrm{md}} = 1$, $w_{\mathrm{mean}} = 1$ and $w_{\mathrm{disp}} = 5$.

**Fig. 9.** Photometric contours, for the 2-integral model (thick curves) and the data (thin curves). A comparison to the previous figure gives an indication for the values of the contours.

Fig. (8) gives an idea of the quality of the fit to the photometry. The upper panel shows the photometry along major and minor axis, together with the model photometry. The lower panel shows the LMD, also along major and minor axis for the gaussian deprojection and as obtained from the model. The correspondence is quite good, especially when taking into account that (1) the LMD was not given as data to the QP program. (2) The model is, as a fit, always a compromise between photometry and kinematics. The model LMD as shown in the figure is thus a derived quantity from the distribution function, as obtained from all available projected moments.

**Fig. 8.** Left: 2-integral model, right: 3-integral model. Upper panel: photometry along minor axis (upper curves) and major axis (lower curves), for the model (full curves) and the data (dash-dot). Lower panel: the LMD along major axis (upper curves) and minor axis (lower curves), for the model (full curves) and the data (dash-dot).

Fig. (9) shows the contours of the model photometry. This offers a more global look on the fit of the photometry than the upper pannel of fig. (8). The isophotes do not quite have a regular ellipsoidal shape, and the fit seems to be better near the centre. This is largely because the gradient in the surface brightness becomes very small in the outer parts, thereby magnifying differences in the location of the contours. Also, the weights used in QP are according to relative errors in the surface brightness, hence absolute errors in the magnitudes, favouring a good fit in the centre. Finally, the compromises due to simultaneously fitting all data tend to favour the fit to the kinematics in the outer parts.

In the left column of fig. (10) the fit to the projected kinematics is shown. The model stays essentially within the bounds dictated by the error bars. The mean rotation on the major axis at about $10''$ is underestimated, but this is probably not an essential defect. It has to do with the very high velocity dispersion around that radius. The rotation at $20''$ is slightly overestimated, but, again, this is unlikely to be a real problem. The main feature here is that the model manages to accommodate the (maybe somewhat too) flat velocity dispersion profile along the minor axis. This was the main difficulty in the modeling.

Our conclusion from this section therefore is that two-integral models cannot be flatly excluded. In the next section we create a 3-integral model and want to investigate whether such a model produces a better fit, especially there where the 2-integral model showed some deficiencies.

## 5. Three-integral models

The algorithm for constructing a 3-integral model is identical to the one previously discussed for a 2-integral model. Of course, here the components involve a thirth integral $I_3$. Working with the Stäckel version of the potential and in spheroidal coordinates, the analytical expression for $I_3$ is known (e.g., Dejonghe & de Zeeuw 1988) and given by

$$I_3 = \frac{1}{2}(L_x^2 + L_y^2) + (\gamma - \alpha)\left[\frac{1}{2}v_z^2 - z^2\frac{G(\lambda) - G(\nu)}{\lambda - \nu}\right], \qquad (19)$$

An expansion in components of Fricke type now reads

$$\begin{aligned}F_{pqn}^{\pm} &= E^p L_z^{2q} 2^{-q}(L_z^2/2 + I_3)^n \qquad \pm L_z \geq 0 \\ &= 0 \qquad \pm L_z < 0,\end{aligned} \qquad (20)$$

**Fig. 10.** Left colum: The 2-integral fit to the kinematic data. Right colum: The 3-integral fit to the kinematic data. Upper panel: major axis, second panel: ray parallel to the major axis at 10", thirth panel: ray parallel to the major axis at 20", lower panel: minor axis. The upper curves are projected dispersions $\sigma_p$, the lower curves are projected rotation, here denoted by $\mu_p$.

where $n$ is an integer, but $p$ and $q$ can be real. Mass densities become

$$\rho^{\pm}_{pqn} = \psi^{p+q+3/2} \varpi^{2q} \sum_{h=0}^{n} (\lambda - \nu)^h \sum_{l=0}^{n-h} (\lambda + \alpha)^l$$

$$\times \sum_{k=0}^{n-h-l} {}_{h,l,k} g^{i,m,j}_{pqn} (G(\nu))^{n-h-l-k} \psi^{h+l+k} \quad (21)$$

with

$${}_{h,l,k} g^{i,m,j}_{pqn} = 2\sqrt{2\pi} \binom{n-h-l}{k}\binom{n-h}{l}\binom{n}{h} \quad (22)$$

$$\times \frac{\Gamma(1/2)\Gamma(k+p+1)\Gamma(q+l+1/2)\Gamma(h+1/2)}{\Gamma(q+p+h+l+k+5/2)}. \quad (23)$$

The moments with respect to $v_\lambda$, $v_\nu$ and $v_\phi$ are

$$\left(\rho\langle v_\lambda^{2i} v_\phi^m v_\nu^{2j}\rangle\right)^{\pm}_{pq} = (\pm)^m \psi^{i+j+\frac{m}{2}+p+q+3/2} \varpi^{2q}$$

$$\times \sum_{h=0}^{n} (\lambda-\nu)^h \sum_{l=0}^{n-h} (\lambda+\alpha)^l$$

$$\times \sum_{k=0}^{n-h-l} {}_{h,l,k} f^{i,m,j}_{pqn} (G(\nu))^{n-h-l-k} \psi^{h+l+k}, \quad (24)$$

with

$${}_{h,l,k} f^{i,m,j}_{pqn} = 2^{i+j+\frac{m}{2}+3/2} \binom{n}{h}\binom{n-h}{l}\binom{n-h-l}{k} \Gamma(i+\frac{1}{2})$$

$$\times \frac{\Gamma(j+h+1/2)\Gamma(q+\frac{m}{2}+l+1/2)\Gamma(k+p+1)}{\Gamma(q+p+h+l+k+i+j+\frac{m}{2}+5/2)}. \quad (25)$$

**Fig. 11.** Photometric contours, for the 3-integral model (thick curves) and the data (thin curves). A comparison to the previous figure gives an indication for the values of the contours.

The 3-integral components are defined such that, when the expressions (21) to (24) are evaluated with $n=0$, they reduce to the corresponding formulae for the 2-integral model.

The physical moments with respect to $v_\varpi$, $v_\phi$ and $v_z$ follow after transformation of (24) with the rotation

$$\begin{pmatrix} v_\omega \\ \text{sign}(z)v_z \end{pmatrix} = \begin{pmatrix} \cos\theta & -\sin\theta \\ \sin\theta & \cos\theta \end{pmatrix} \begin{pmatrix} v_\lambda \\ \text{sign}(\gamma-\alpha)v_\nu \end{pmatrix}, \quad (26)$$

where

$$\cos\theta = \sqrt{\frac{(\nu+\alpha)(\lambda+\gamma)}{(\alpha-\gamma)(\lambda-\nu)}}, \qquad \sin\theta = \sqrt{\frac{(\lambda+\alpha)(\nu+\gamma)}{(\gamma-\alpha)(\lambda-\nu)}} \quad (27)$$

as in Dejonghe & de Zeeuw (1988). The 3-integral distribution function has the form

$$\sum_{p,q,n} \left( c^e_{pqn} F^e_{pqn}(E, L_z, I_3) + c^-_{pqn} F^-_{pqn}(E, L_z, I_3) \right). \quad (28)$$

| $p$ | 2, 3, 4, 5, 6, 7, 9, 10, 15 |
|---|---|
| $q$ | 0, 0.1, 0.25, 0.5, 0.75, 1, 2, 3, 4, 5, 7, 9 |
| $n$ | 0, 1, 2, 3, 4, 5 |

**Table 5.** The range for the parameters in the Fricke components, $p$ an $q$ are used in both models, $n$ only in the 3-integral model.

Table (5) lists the values for the parameters $p$, $q$ and $n$ which the program could use to choose a Fricke component as in (20). The 3-integral distribution function is made with 62 components. We expect that a 3-integral model gives at least as good a fit as a 2-integral model. It is constructive to pay special attention to the curves that were less well fitted by a 2-integral-distribution function.

In fig. (8) we can compare the photometry for both models. The 3-integral (right) model gives smoother photometry and LMD profiles on the whole, especially along the major axis (lower curves in upper panel and upper curves in lower panel). This is most apparent when comparing the model LMD and the deprojected LMD. The contours for the 3-integral model photometry in fig. (11) show isophotes that are smoother and more ellipsoidal than in fig. (9). Again the fit is better near the centre.

The model and the data for the projected kinematics are in good agreement, as can be seen in the right column of fig. (10) The mean rotation at $10''$ is still overestimated, but at $20''$ the rotation curve has flattened and stays mainly within the error-bars. The mean rotation along the major axis tends to be a bit too low far from the centre, but both mean rotation and velocity dispersion are definitely smoother and agree better with the data than the 2-integral model. Also the flat velocity dispersion profile along the minor axis is better fitted with the 3-integral model than with the 2-integral model. We therefore conclude that this particular 3-integral model produces a better fit. Judging from this case, 3-integral clearly constitute an additional value over 2-integral models, and it is not unreasonable to expect that this wil be the case for many galaxies of this kind.

## 6. Results

### 6.1. The mass-to-light ratio

It is well-known that the presence of kinematical information of high quality is essential for determining masses. Unfortunately, kinematical information for elliptical galaxies does not constrain the mass as tightly for ellipticals as it does for spirals. The reason for this is clear: the orbital structure of ellipticals is, in principle, fairly arbitrary, while there are very good reasons for assuming circular orbits for the stars (and the gas) in spirals.

It is possible to base the above statements on a firm theoretical footing. One can show that, technically speaking, the distribution function of a spherical anisotropic galaxy is uniquely determined by its augmented mass density, which is a function $\tilde{\rho}(\psi, r)$ that depends on the radius and on the potential function (Dejonghe 1986). It is thus a function of 2 variables, in agreement with the number of variables in the anisotropic distribution function. However, the ordinary mass density $\rho(r)$ only depends on $r$, and $\rho(r) = \tilde{\rho}(\psi(r), r)$. Hence, there are many distribution functions possible for a given mass density (not all of them will be positive though). Therefore, attention has shifted to the interpretation of line profiles (Dejonghe 1987, Gerhard 1993). In fact, a theorem (Dejonghe & Merritt 1992) asserts that the distribution function is completely determined if the potential and all the line profiles are known.

The situation is considerably different for axisymmetric galaxies. There, also, the 2-integral distribution function follows uniquely from the augmented mass density which is now a function $\tilde{\rho}(\psi, \varpi)$. In contrast to the spherical case, however, this function is uniquely determined by the mass density $\rho(\varpi, z)$. Hence, the degeneracy in the 2-integral orbital structure is essentially absent, and kinematical data will constrain the potential much better than in the spherical case.

| R (kpc) | R (arcsec) | M ($10^{11} M_\odot$) |
|---------|------------|-----------------------|
| 2       | 17         | 0.4                   |
| 4       | 34         | 0.8                   |
| 6       | 51         | 1.2                   |
| 8       | 68         | 1.5                   |
| 10      | 85         | 1.8                   |
| 12      | 102        | 2.0                   |
| 14      | 119        | 2.2                   |
| 16      | 136        | 2.3                   |
| 18      | 153        | 2.4                   |
| 20      | 170        | 2.6                   |

**Table 6.** The cumulative mass function, indicating the total mass inside a sphere with specified radius.

This is confirmed by our numerical experience with the models for NGC 4697. The mass-to-light ratio is a reasonably well determined $\Upsilon_V = 3.5 h_{50}$ at a distance of $24 h_{50}^{-1}$ Mpc, with an allowance of a few tenths. Taking $B - V = .95$ (Faber et al. 1989), this translates into $\Upsilon_B = 4.8 h_{50}$, in very good agreement with Binney et al. (1990). Finally, in table (6) we list a few cumulative masses, as calculated from the zeroth order harmonic of the total matter density.

6.2. The distribution function

An essential feature in our modeling is the inclusion of a distribution function. It turns out that the distribution functions we have constructed can be suitably expressed in units of $L_\odot \text{pc}^{-3}(100 \text{km/s})^{-3}$. The values so obtained can then easily be multiplied by the mass-to-light ratio $\Upsilon$ in order to obtain phase space densities in units of $M_\odot \text{pc}^{-3}(100 \text{km/s})^{-3}$.

The 3 integrals $E$, $L_z$ and $I_3$ are not very intuitive. We will replace them by 3 others, which are closer to the morphology of a typical orbit.

In the 2-integral case, orbital families, labeled by $E$ and $L_z$ fill a torus-like structure in configuration space. The torus intersects the equatorial plane in a ring, with inner radius $R_-$ and outer radius $R_+$. These we call the turning points, and there is a unique relation between them and the classical integrals. The 2 turning points are just another set of integrals, which happen to be somewhat more intuitive.

In the 3-integral oblate Stäckel case, such as we consider, the orbits are called short axis tubes, and are labeled by $E$, $L_z$ and $I_3$. In configuration space they are bounded by 2 spheroids of constant $\lambda$ and one hyperbola (of 2 sheets) of constant $\nu$. The intersection of the ellipses with the equatorial plane is a ring, with inner radius $R_-$ and outer radius $R_+$. The intersection of the outer ellipse with the hyperbola is a circle with radius $R$ and height $Z$. We found it more easy to substitute the coordinate $\nu$ with the pair $(R, Z)$. The set $(R_-, R_+, (R, Z))$ we call the turning points, and there is a unique relation between them and the 3 integrals (de Zeeuw 1985a).

The turning points $R_-$, $R_+$ and $(R, Z)$ are positive numbers by definition. However, we need also to take account of the sense of rotation of the stars, i.e. of the sign of $L_z$. This we do by assigning the sign of $L_z$ to $R_-$.

At present, no distribution functions for NGC 4697 existed in literature that can be compared to ours in a quantitative way. Therefore, we will first present our distribution functions by comparing the 2-integral and 3-integral versions. This is only meaningful for orbits in the equatorial plane, which exist in both cases. In general, these orbits fill a ring with inner radius $R_-$ and outer radius $R_+$. All other orbits may be disjunct to both models.

Fig. (12) shows this comparison. The upper panel represents the 2-integral distribution function for orbits up to $R_{\max} = 5$kpc and the lower panel shows the corresponding 3-integral distribution function. For $R_+$ between 0 and 3kpc, both distribution functions have a very similar structure. The extra freedom that the 3-integral model offers, seems to have mainly effect on the shape of the distribution function at larger radii. By and large, the 3-integral has the less complicated structure. The prevalence of orbits that are close to circular, which is always present in flattened 2-integral models (Dejonghe 1986, Dehnen & Gerhard 1993) is clearly not so large in the 3-integral model. The additional freedom in the third integral is used to produce a flattened figure by populating orbits with small $I_3$, which remain close to the equatorial plane. In general, the 3-integral model seems to distribute the orbits more evenly over $L_z$. The orbits with a fixed pericentre in the 3-integral distribution function, are distributed more smoothly than in the 2-integral distribution function, as can be seen in the lower panel. In both models though, the orbits with apocentrum = 0 kpc are clearly present. Here again, we notice the dominating number of orbits with negative angular momentum.

The cut of the 3-integral distribution in the upper panel of fig. (13) shows the distribution of stars for orbits that stay in the equatorial plane, which can be compared to the cut in the lower panel, representing orbits that reach the hyperbola passing through $(R, Z) = (5, 5)$ kpc. We clearly notice that orbits that reach fairly large heights, or that have larger values for $I_3$ are less dense populated. Moreover, the distribution becomes more isotropic, especially near the centre.

**Fig. 13.** 3-integral distribution functions for a constant $\nu$, which represents orbits delimited in height by a hyperbola. In the lower panel, this hyperbola passes through $(\varpi, z) = (5,5)$ kpc, while in the upper panel orbits remain confined to the equatorial plane. The labels are in $\log(L_\odot \text{pc}^{-3}(100\text{km/s})^{-3})$.

**Fig. 12.** The distribution function for the 2-integral (upper panel) and the 3-integral model (middle panel) for orbits in the equatorial plane. Pericentre and apocentre are denoted by $R_-$ and $R_+$. The distribution of orbits with fixed pericentre ($R_+ =$3kpc) in the 2-integral model (full lines) and in the 3-integral model (dash-dot lines) in the lower pannel. The labels are in $\log(L_\odot \text{pc}^{-3}(100\text{km/s})^{-3})$.

Rather than using the ratio's $a$ en $b$ as in Dehnen & Gerhard (1993) to quantify the importance of a third integral, we will illustrate the dependence on $I_3$ in the most direct way possible: by simple inspection. Fig. (14) shows some cuts of the 3-integral distribution function with a plane of constant $L_z$, three of which have negative $L_z$, since our choice of the axes was such that most orbits have negative $L_z$. In the case of a 2-integral distribution function, we would see vertical contours. Clearly, the dependence on $I_3$ is significant, and adds additional structure to the distribution function. The morphology of integral space is such that with increasing $|L_z|$, the intervals in $E$ and $I_3$ that correspond to real orbits become smaller (see figure 1 in Dejonghe & de Zeeuw 1988). The maximum density of orbits in these planes is found for large values of the binding energy $E$ and small values of $I_3$; this is true for all 4 panels though the absolute number densities are decreasing. In fig. (15), orbits that are confined to a meridional plane are presented for both models. The 2-integral model needs no further comment, of course, but the 3-integral model shows, apart from an obvious preponderance of tightly bound orbits, a tendency to populate elliptic orbits over the pole, not unlike a polar ring!

**Fig. 14.** Contours for the 3-integral distribution function in a few planes of constant $L_z$. The distribution functionis given as function of $(E, 4EI_3)$, thereby avoiding large values in the second coordinate.The labels are in $\log (L_\odot \text{pc}^{-3}(100\text{km/s})^{-3})$.

**Fig. 15.** Meridional orbits (i.e. with $L_z = 0$) in the 2-integral model (left panel) and the 3-integral model (right panel). The values for the contours are -4.9,-4.5,-4,-3.5 and -3 $\log(L_\odot \text{pc}^{-3}(100\text{km/s})^{-3})$.

### 6.3. Moments

The anisotropy of the velocity dispersion tensor, as derived from the 2-integral model, is shown in fig. (16). From the upper panel we see that by and large $\sigma_\phi > \sigma_\varpi$, which is the same conclusion as obtained by Binney *et al.* (1990). This is especially true around $\varpi = 2$ kpc, which follows, again, from the very high projected velocity dispersions in that region. Only around the minor axis is $\sigma_\varpi > \sigma_\phi$, which is likely a result from the high velocity dispersion along the minor axis. As indicated, on the minor axis itself, $\sigma_\varpi = \sigma_\phi$, as should because of symmetry. The Satoh parameter $k$ on the other hand clearly is not a constant. It is in the same range as quoted by Binney *et al.* (1990), but on the locus where $\sigma_\varpi = \sigma_\phi$ in the upper panel, we find a ragged contour where $k$ is discontinuous and infinite. To the left of it, $k$ is imaginary, according to definition (2), and $k$ is undetermined on the minor axis. Therefore, we adopted a slightly different definition:

$$k = \frac{\mu_\phi}{\sqrt{|\langle v_\phi^2 \rangle - \sigma_\varpi^2|}} \times \text{sign}(\langle v_\phi^2 \rangle - \sigma_\varpi^2) \qquad (29)$$

Note also that $10''$ corresponds roughly to 1 kpc: beyond 5 kpc the kinematic features are not very constrained by the data, which at these distances from the centre consist only of photometry.

Fig. (17) conveys an impression of the degree of anisotropy in the velocity dispersion tensor for the 3-integral model, which is to be compared to the analogous fig. (16) for the 2-integral model. In this case, one also has to consider the difference between $\sigma_\varpi$ and $\sigma_z$. It is immediately apparent that the 3-integral model produces more constant fields $\sigma_\phi - \sigma_\varpi$ and $k$, which is to be expected since this is mostly a feature determined by the phase space structure in $L_z$, which is more complicated for the 2-integral model.

In general, the velocity dispersion tensor is only mildly anisotropic. In particular, the upper panel shows that there are only minor differences between $\sigma_\varpi$ and $\sigma_z$. This is not surprising since it was already shown that a 2-integral distribution function produces an acceptable model for the data. More importantly, $\sigma_\varpi \geq \sigma_z$ in the region that is constrained by kinematic data, as can be expected from a flattened galaxy. Moreover, the differences between the radial and tangential dispersion in the middle panel are of the same order of magnitude as in the upper panel and have somewhat decreased in comparison with the 2-integral model. This also follows from the bottom panel (the parameter $k$), which is qualitatively similar to the 2-integral analog.

**Fig. 16.** Contours for $\sigma_\phi - \sigma_\varpi$ (upper panel) and the Satoh parameter $k$ (lower panel) for the 2-integral model. The contours in the upper panel are labeled in km/s.

In fig. (18) we see the spatial dispersions along major axis. From top to bottom: $\sigma_\varpi$, $\sigma_z$ and $\sigma_\phi$ for the 2-integral model in dash-dot lines and for the 3-integral model in full lines. The radial (and the azimuthal dispersion) of 2-integral distribution function attain their highest values (175 km/s) near the centre, which correspond well to the observed central projected value of 180 km/s. The azimuthal dispersion has a high maximum of about 220 km/s at 1.3 kpc, clearly associated with the peak value at $10''$ in the projected dispersion. At larger radius, all 2-integral dispersion profiles decrease along the major axis.

The 3-integral distribution function on the other hand has a higher central velocity dispersion (190 km/s). The azimuthal dispersion again follows closely the projected dispersion profile. While $\sigma_\varpi$ contributes to the projected central velocity dispersion, the $\sigma_z$ profile is not detectable in line-of-sight kinematic data. Hence, it is probably weakly constrained by the data.

Fig. 19 shows the spatial dispersions along minor axis. The 2-integral model tries hard to keep up with the very flat projected velocity dispersion: the intrinsic dispersion rises in the interval that is constrained by the data. The 3-integral model however is not as constrained by the photometry as the 2-integral model. The radial velocity dispersion in that case is very high (close to 200 km/s) and stays fairly flat. The $\sigma_z$ on the other hand is only 70% of $\sigma_r$. The higher 3-integral dispersions are not quite surprising, since on the whole the 3-integral model produces a better fit with its higher projected dispersions on the minor axis.

**Fig. 17.** Contours for $\sigma_\varpi - \sigma_z$, (upper panel), $\sigma_\phi - \sigma_\varpi$ (middle panel) and the Satoh parameter $k$ (lower panel) for the 3-integral model.

Fig. (20) shows the mean velocity field for the 2-integral model in the upper panel and the 3-integral model in the lower panel. Both fields are fairly regular, as one might have expected. Close to the rotation axis the mean rotation is almost zero, and can therefore as well be somewhat negative since the number densities of clockwise rotating orbits and counter clockwise rotating orbits need not exactly cancel. This explains the presence of a region with negative velocities, which is, not unexpectedly, larger in the 3-integral model. The structure present in the projected velocity profiles along the major axis (fig. (10) are also present here, including the location of the peak velocity. This peak is closer towards the centre in the 3-integral model.

For the second order moments, there are a few remarks that hold for the radial dispersion in fig. (21) as well as for the vertical dispersion in fig. (22). In the 2-integral model, the radial and the vertical dispersion are the same. For these contours, the peak values are located near the centre. The contours for the highest dispersions for the 3-integral model have moved away from the centre. This indicates that the extra parameters offered by the 3-integral distribution function are well used for modeling the data. In fig. (23) we see that the 2-integral model has high values for the azimuthal dispersion in more extended regions than the 3-integral model. The 3-integral azimuthal dispersion field is more regular than the 2-integral one.

**Fig. 18.** The second order moments along major axis, for the 2-integral model (dash-dot lines) and for the 3-integral model (full lines). Upper panel: $\sigma_\varpi$. Middle panel: $\sigma_z$. Lower panel: $\sigma_\phi$.

## 7. Predictions

### 7.1. Slit positions

As already mentioned in section 2, the data obtained at $5''$ parallel to major axis and at 45° intermediate axis were not used for the modeling. We will compare the predictions from our models at these slit positions with the observed kinematics. This gives an indication of the amount of freedom that is still available for the model when using kinematical data on four slit-positions.

The upper panels of fig (24) show that the dispersion at $5''$ from the major axis behaves very similar to that on the major axis (see fig(10)), and hence the fit is good. The line-of-sight velocities of the 2-integral model however show already larger deviations, especially at small radii. This behaviour could be expected, since the 2-integral model could not quite keep up with the steeply rising rotation curve along the major axis, as can be seen in fig(10), and continues this trend here. On the contrary, the 3-integral model reproduces the rotation curve well (fig(10)), and does so also at $5''$. We conclude that in this case, the slit at $5''$ offset from the major axis does not seem to offer substantial additional information.

The lower panels of fig(24) show that on the 45° intermediate exis the predicted behaviour is a fair fit to the data. The dispersion is in places somewhat too high and too low, but the

**Fig. 24.** Left colum: The 2-integral model in comparison with the data at 5″ parallel to the major axis (upper panel) and at 45° intermediate axis (lower panel).

**Fig. 19.** The second order moments along minor axis, for the 2-integral model (dash-dot lines) and for the 3-integral model (full lines). Upper panel : $\sigma_\varpi$. Lower panel: $\sigma_z$.

**Fig. 20.** Contours for the mean velocity for the 2-integral distribution function (upper panel) and the 3-integral distribution function (lower panel).

### 7.2. Line profiles

line-of-sight velocity data are fully consistent with the model. Clearly the intermediate slit position adds non-redundant data, at least for the dispersions.

Since this paper is partly based on data which are taken quite some time ago, it proved impossible to retrieve significantly more information out of the spectra regarding the shape of the line profiles, that was not already contained in the first 2 moments. On the other hand, it is always possible to calculate line profiles from the models, and fig. (25) shows two of them. Table (7) gives the thirth and fourth order Gauss-Hermite coefficients, which characterise the deviations from Gaussians (Gerhard 1993, van der Marel & Franx 1993, Magorrian & Binney 1995).

**Fig. 21.** Contours for the radial dispersion $\sigma_\varpi$ for the 2-integral distribution function (upper panel) and the 3-integral distribution function (lower panel).

**Fig. 22.** Contours for the vertical dispersion $\sigma_z$ for the 2-integral distribution function (upper panel) and the 3-integral distribution function (lower panel).

On the minor axis, at $10''$, there is little or no difference between the models. Both profiles very much look like symmetric gaussians ($h_3 = 0$). The 3-integral line profile in this point has a more flattened top, but the deviation from a Gaussian is considerably smaller than the line profile of the 2-integral model.

On the other hand, on the major axis, there seems to be a detectable difference between both models. Judging from the

**Fig. 23.** Contours for the azimuthal dispersion for the 2-integral distribution function (upper panel) and the 3-integral distribution function (lower panel).

$h_3$'s, the line-profile for the 2-integral model is skewer than that for the 3-integral model. This skewness cannot result from projection along line of sight, but should be present in the model (see also van der Marel & Franx (1993)). In any case, both line profiles are clearly skewed, with the broader wing the retrograde one, as was also noted by Bender et al. (1994). The small bump in the 2-integral model is probably not very significant.

Finally, the most obvious difference is the value of the mean projected rotation as can be seen from figure (10) : the global fit has produced a mean rotation for the 3-integral model of about 85 km/s, while the 2-integral model rotates at 120 km/s, where the model actually shows a bump in the rotation curve.

| position | model | $h_3$ | $h_4$ |
|---|---|---|---|
| major axis at $30''$ | 2I | -0.237 | -0.073 |
|  | 3I | -0.042 | -0.017 |
| minor axis at $10''$ | 2I | 0 | 0.039 |
|  | 3I | 0 | -0.015 |

**Table 7.** The Gauss-Hermite coefficients for a few selected line-profiles.

## 8. Conclusions

The main goal of this paper is the construction of 2-integral and 3-integral dynamical models for the E4 elliptical NGC 4697. To achieve this, we proceeded along the following lines:

– We acquired high quality photometry (NTT and HST), and confirmed the presence of a disk (Carter 1987, Peletier 1990). In addition, we discovered a nuclear dust lane at $3.4''$ from the centre.
– We constructed the deprojected luminous mass density, using 2 well-known methods. We conclude that the gaussian

**Fig. 25.** Line profiles calculated from the models, on the minor axis at $10''$ (lower panel) and on the major axis at $30''$ (upper panel), both normalized to 1. Full lines are for the 3-integral distribution function and dash-dot lines are for the 2-integral distribution function.

expansion method, if applicable, produces very similar results as the Lucy deprojection, but has the advantage of yielding a smooth mass density in analytical form. We use a potential that follows from a constant mass-to-light ratio.
- We use the kinematical data of Binney et al. (1990), supplemented with our own data on the major axis. Especially in the central regions, NGC 4697 appears to be a bulge-like rapid rotator.
- Our models were constructed using quadratic programming. This method works well, and allows the construction of 2-integral and 3-integral models alike. The models are analytical. The third integral is taken from the exact third integral in the Stäckel fit to the potential. Hence it is analytical and simple, but approximate.

We further note the following results more in particular:
- NGC 4697 has a $(M/L)_B = 4.8 h_{50}$, wich is moreover reasonably well determined. Judging from previous experience (Bertin et al. 1994), the assumption of axisymmetry is much more constraining on the mass-to-light ratio than the assumption of anisotropic sphericity, whether or not line profiles are present. This seems to be the case at least for very flattened ellipticals as the one we studied here.
- The 2-integral distribution function has a more complicated structure (as a function of $L_z$) than the 3-integral model. This we interpret as a result of the (artificial) independence of $I_3$ for the 2-integral model, which renders the problem of finding a suitable distribution function in general more difficult. The dependence on the third integral is mostly used by the 3-integral model in order to produce flattening by the well-known mechanism $\sigma_\varpi > \sigma_z$, which means that orbits remain closer to the equatorial plane if judged by their radial excursions. As a corrollary, the 2-integral model contains more orbits that are close to circular than the 3-integral model.
- We noticed that the Satoh-parameter $k$ does not remain constant in our 2-integral model. We adopted a slightly different definition in order to cope with the fact that in regions close to the rotation axis $\sigma_\varpi > \sigma_\phi$. This suggests that the assumption of a constant (and positive) $k$ might be responsible for a 2-integral model that is not quite satisfying (Binney et al., 1990).
- Both models show an anisotropy where $\sigma_\varpi > \sigma_z$ in the region where kinematical data are available, as can be expected for a flattened galaxy. The velocity dispersion fields in the 3-integral model are more constant as a function of radius than the corresponding fields in the 2-integral model, which, in turn, show more complicated structure.
- Not surprisingly, the extra freedom that the 3-integral model offers is used for a better fit to the peculiarities in the velocity and dispersion profiles. Nevertheless, the first and second order moments for the 3-integral distribution function are the smoother. This is also the case for the calculated line profiles.
- In the presence of kinematical data along major axis, minor axis, and 3 slit positions $\parallel$ to the major axis, it turns out that data along the $45°$ intermediate axis offers non-redundant information.

*Acknowledgements.* WWZ acknowledges the support of the Austrian *Fonds zur Förderung der wissenschaftlichen Forschung* in the framework of an Erwin–Schrödinger–Fellowship (project J00934–PHY).